\documentclass[a4paper]{panl}
\usepackage{cite}
\usepackage{graphicx}
\usepackage{amssymb}
\usepackage{amsfonts}
\usepackage{amsmath}
\usepackage{rotating}

\originalTeX
\begin{document}

\title{Methodology for measuring gluon jet fraction and characteristics 
   of quark and gluon jets for hadron-hadron collisions}
\maketitle
\authors{ S.\,Shulha$^{a,b,}$\footnote{E-mail: siarhei.shulha@cern.ch}
D.\,Budkouski$^{a,c,}$\footnote{E-mail: dzmitry.budkouski@cern.ch} }
\from{$^{a}$\,Joint Institute for Nuclear Research, Dubna}
\vspace{-3mm}
\from{$^{b}$\,F.Skorina Gomel State University, Gomel, Belarus}
\vspace{-3mm}
\from{$^{c}$\,Institute for Nuclear Problems of Belorusian State University, Minsk, Belarus}

\begin{abstract}
  The discriminators of quark-gluon jets developed for $pp$ collisions at the LHC 
can be used to measure the gluon jet fraction in a jet sample.
  It allows to measure various characteristics of the quark and gluon jets.
  The methodology of such measurements for modern hadron colliders is discussed.
\end{abstract}
\vspace*{6pt}

\noindent

\label{sec:intro}
\section*{ \bf INTRODUCTION }
   The gluon- and quark-initiated jets ($g$- and $q$-jets) have essentially different properties:
charged particle multiplicity is larger inside $g$-jets than inside $q$-jets,
$g$-jets are less collimated than $q$-jets, energy of $g$-jet is more smooth distributed between
hadrons, while most of the $q$-jet energy is concentrated in several leading hadrons.
   The properties of $q$- and $g$-jets were first studied in $ e^+e^- $ 
processes which has small number of jets per event, thus it is possible to separate 
$q$- and $g$-jets according to the channel signature~\cite{Dremin:2001}.
  In hadron-hadron collisions, multijet events occur more frequently. 
  For the recognition of $q$- and $g$-jets in hadron-hadron collisions, 
special methods are developed based on discriminant analysis. 
  Jet physical characteristics are used to build another one, 
which is the most sensitive to the jet type, and it is called a discriminator ($D$) 
of q/g-jets~\cite{QGL8TEV:2013, QGL13TEV:2017}.
   
  Discriminators of $q/g$-jets are mainly used to select channels involving jets. 
  Choosing jets that are above a certain operating working point $D > D_0$, 
one can suppress true or false $q/g$-jets with the required efficiency. 
  The quality of the discriminator, which is characterized by the difference 
between the $D$-distributions for the $q$- and $g$-jets, 
strongly affects the purity of the channel selection and the number of selected events.

  Another class of measurements, in which the discriminator of $q/g$-jets can be applied, 
is related to measurements of fraction of $q/g$-jets in the jet sample ($\alpha^{q/g}$). 
  Measurement of $\alpha^{q/g}$ can be performed using method with a fixed operating working point.
  But more promising is the method without using a working point by
fitting the measured distribution over the whole $D$-region.
  In the latter case, the statistical uncertainty of measurement is based on the entire jet sample.
  Therefore, a sufficiently high measurement accuracy can be achieved without 
imposing stringent requirements on the quality of the $D$-discriminator, 
if the sample of jets is sufficiently large. 
  This is especially important for tasks where the quality of the $D$-discriminator 
is limited by the physical nature of the recognized objects.
  The hadron collider provides the ability to produce a large number of jets, 
which allows to perform precision measurements of $\alpha^{q/g}$.

  The high accuracy of measurement of the $q/g$-jet fractions, in turn, imposes new requirements
to the methods of measuring the characteristics of $q/g$-jets.
  Since the characteristics of $q$- and $g$-jets differ, the characteristics of jet sample 
are mainly determined by the ratio of $q/g$-jet fractions.
  However, there are factor of the next order of accuracy that also affect 
the characteristics of jet sample: it is dependence of the properties of $q/g$-jets 
on jet sample, i.e. on kinematics and environment in which the selected jets are formed 
and collected.

   Until now, measurements of the characteristics of $q/g$-jets
(distributions over the multiplicity of hadrons in jet and 
the mean multiplicity in jet) at hadron colliders were performed 
using the generator values of the $q/g$-jet fractions in jet sample~\cite{CDF:2005, ATLAS:2016}.
   Precision measurements of  $q/g$-jet fractions open up new possibilities 
in measurement of $q/g$-jets characteristics.

\label{sec:Gfract}
\section*{ \bf  MEASUREMENT OF GLUON JET FRACTION }

  The fraction of $g$-jets in jet sample can be determined by fitting the measured
normalized $D$-distribution of jets, $H^{DAT}(D)$, with a one-parameter function:
\begin{equation}\label{eqs:QGLFitFunction}
     H^{DAT}(D) \sim \alpha^g H^{g\,MC}(D) + (1 - \alpha^g) H^{q\,MC}(D).
\end{equation}
  The symbol $\sim$ means fitting
of the measured  distribution 
of reconstructed jet sample on the left side of equation
with the function on the right side, 
which is a linear combination of the Monte Carlo (MC) 
distributions of $q/g$-jets, $H^{q/g\,MC}(D)$, 
with an unknown $g$-jet fraction $\alpha^g$.
  Model-dependent distributions $H^{q/g\,MC}(D)$  are determined
from MC sample of reconstructed jets (with full simulation of detector response), 
which type ($q$-jet or $g$-jet) is found 
by matching the 
reconstructed and generator jets.
  In Eq.~(\ref{eqs:QGLFitFunction}) $q$-jets are jets with all possible 
quark flavors, including reconstructed jets with misidentified flavour, 
for which there were no corresponding jets found among the generator jets.
  
   For verification of the $\alpha^g$ uncertainty, which is provided by method of fitting 
the Eq.~(\ref{eqs:QGLFitFunction}), method of "random cloning of the experiment" can be used:
for all histograms involved in the fit, random clones of histograms  
(which are the histograms with the number of entries equal to the number of entries 
in the original histograms and randomly distributed according to the original histograms) 
are constructed, and the fitting procedure is repeated.
   Variance of the found $\alpha^g$ values is an estimate of the standard deviation 
of the measured average $g$-fraction, and includes statistical uncertainty
and systematic uncertainty of the fitting procedure~(\ref{eqs:QGLFitFunction}).

   Stable result of fit procedure Eq.~(\ref{eqs:QGLFitFunction}) can be obtained 
using the method of weighted least squares, in which the following value is a subject for minimization:
\begin{equation}\label{eqs:QGLwlsMethod-V}
 V = \sum_{D}  \Big[\frac{Y(D)}{w(D)}\Big]^2 = \min,
\end{equation}
where
\begin{equation}\label{eqs:QGLwlsMethod-Yi}
   Y(D) = H^{DAT}(D) - \alpha^g H^{g\,MC}(D) - (1-\alpha^g)H^{q\,MC}(D).
\end{equation}
   In Eq.~(\ref{eqs:QGLwlsMethod-V}), the summation is performed over the bins of $Y(D)$ histogram.
   Weight, $w(D)$, can be chosen as the standard deviation of the content of the bin $D$
for the measured histogram $H^{DAT}(D)$, assuming that the model distributions $H^{q/g\,MC}(D)$ are known 
with good accuracy: $w(D) = \Delta H^{DAT}(D)$.

\label{sec:QGcharackteristics}
\section*{\bf MEASUREMENT OF CHARACTERISTICS OF QUARK AND GLUON JETS }
  
  Suppose there are two jet samples with the measured characteristics $X_k$ ($k=1,2$).
  In this section, the characteristic of the jet sample $X_k$ stands for the normalized distribution 
of jets over some jet parameter $y$: $X_k\equiv X_k(y)$.
  Generalization of conclusions and formulas presented below to the case of raw moments 
of distribution over $y$ as a jet sample characteristic $X_k$ can be performed trivially.
  
  Let $\alpha_k^g$ be g-jet fraction in $k$th jet sample, which is measured according 
to Eq.~(\ref{eqs:QGLFitFunction}). 
  Then there is a system of two equations for unknown characteristics of $q/g$-jet subsamples, $X^{q/g}$:
\begin{equation}\label{eqs:DDR}
\begin{split}
   X_1 &= \alpha_1^g X^g + (1 - \alpha_1^g) X^q, \\
   X_2 &= \alpha_2^g X^g + (1 - \alpha_2^g) X^q.
\end{split}
\end{equation}
  The solution of system of Eqs.~(\ref{eqs:DDR}) has the form:
\begin{equation}\label{eqs:DDRSolution}
\begin{split}
   X^q = \frac{\alpha_2^g X_1 - \alpha_1^g X_2}{\alpha_2^g - \alpha_1^g}, \,\,\,\,\,\,\,\,
   X^g = \frac{(1-\alpha_1^g) X_2 - (1-\alpha_2^g) X_1}{\alpha_2^g - \alpha_1^g}.
\end{split}
\end{equation}
  
  It is assumed in Eqs.~(\ref{eqs:DDR}) -~(\ref{eqs:DDRSolution})
that $q$- and $g$-jet samples have universal characteristics,
i.e.  $q/g$-jet characteristics  $y$ and $X \equiv X(y)$ are independent on the jet sample they belong to.
  In practice, it is satisfied only approximately. 
  Deviation of characteristic of jets with a particular flavor between different jet samples 
is defined here as "jet flavour non-universality" (JFNU).
  Given JFNU, Eqs.~(\ref{eqs:DDR}) take the form:
\begin{equation}\label{eqs:QGL-FIT-withFNU}
\begin{split}
   X_1 &= \alpha_1^g X_1^g + (1 - \alpha_1^g) X_1^q, \\
   X_2 &= \alpha_2^g X_2^g + (1 - \alpha_2^g) X_2^q.
\end{split}
\end{equation}

  To describe JFNU quantitatively, one can introduce the JFNU measure for $f$-jets, $\Delta X^f$,
and the "universal" characteristics of $f$-jets, $X^f$, 
for which values averaged over two jet samples can be chosen:
\begin{equation}\label{eqs:JFNU-universal}
\begin{split}
  \Delta X^f &\equiv \,\, X_2^f - X_1^f,  \\
   X^f       \equiv& \,\, \rho_1^f X_1^f + \rho_2^f X_2^f,
\end{split}
\end{equation}
  The JFNU measure, $\Delta X^f$, is found by MC simulation, and it characterizes
physical and kinematic differences between the two jet samples.
  The following notation is used in Eqs.~(\ref{eqs:JFNU-universal}):
\begin{equation}\label{eqs:JFNU-universal-where}
\begin{split}
  X^f &\equiv \frac{ n_1^f(y) + n_2^f(y) }{ N^f }, \,\,\, 
   N^f\equiv ( N_1^f + N_2^f ),\,\,   N_k^f\equiv \sum_y n_k^f(y),\,\, k=1,2,\\
  X_k^f &\equiv \frac{n_k^f(y)}{N_k^f},\,\,\,\,\,  
  \rho_k^f \equiv \frac{N_k^f}{N^f},\,\,  \rho_1^f + \rho_2^f = 1,\\
  f &\equiv q,g, \,\,\,\, q=u,d,s,c,b,x. \\
\end{split}
\end{equation}
  Here $n_k^f(y)$ is $y$-distribution of $N_k^f$ jets with the flavor $f$, 
collected in the $k$th jet sample.
  Recall that the dependence of the characteristics on the $y$ is omitted:
$X^f\equiv X^f(y)$, $X_k^f\equiv X_k^f(y)$.
  
  From Eqs.~(\ref{eqs:JFNU-universal}) one can find "non-universal" 
characteristics for two jet samples:
\begin{equation}\label{eqs:JFNU-nonuniversal}
\begin{split}
  X_1^f &= X^f -\rho_2^f \Delta X^f,  \\
  X_2^f &= X^f +\rho_1^f \Delta X^f.
\end{split}
\end{equation}

   Substituting the Eqs.~(\ref{eqs:JFNU-nonuniversal}) into the Eqs.~(\ref{eqs:QGL-FIT-withFNU}), 
a system of equations for the "universal" $q/g$-jet characteristics $X^{q/g}$ can be written:
\begin{equation}\label{eqs:DDR-withJFNU}
\begin{split}
   \tilde{X}_1 &= \alpha_1^g X^g + (1 - \alpha_1^g) X^q, \\
   \tilde{X}_2 &= \alpha_2^g X^g + (1 - \alpha_2^g) X^q,
\end{split}
\end{equation}
 The left parts are the JFNU-corrected measured characteristics, which have the form:
\begin{equation}\label{eqs:QGL-FIT-withFNU-tilde-where1}
\begin{split}
 \tilde{X}_1 &\equiv X_1 + \gamma^{\rm DAT}\Delta X^{\rm JFNU},   \\
 \tilde{X}_2 &\equiv X_2 - \Delta X^{\rm JFNU}.
\end{split}
\end{equation}
  Here the following notation is used:
\begin{equation}\label{eqs:QGL-FIT-withFNU-tilde-where2}
\begin{split}
 \Delta X^{\rm JFNU} &\equiv \beta^q \Delta X^q + \beta^g \Delta X^g, \\
 \beta^f \equiv \frac{\alpha_1^f\alpha_2^f}{\alpha_1^f + \gamma^{\rm DAT} \alpha_2^f},\,\,\, 
  \gamma^{\rm DAT} &\equiv \frac{N_2}{N_1}, \,\,\, f=q,g,\,\, \alpha_k^q = 1 - \alpha_k^g,
\end{split}
\end{equation}
where $N_k \equiv N_k^q + N_k^g $ is the number of jets in $k$th jet sample.
  Thus, the JFNU corrections for the measured characteristics 
are determined by the following: 
measured $g$-jet fractions, $\alpha_k^g$,
ratio of numbers of jets in the two jet samples, $\gamma^{\rm DAT}$, and
model-dependent JFNU measures, $\Delta X^f$.

  Solving the system of Eqs.~(\ref{eqs:DDR-withJFNU}), one can obtain
the "universal" characteristics of $q/g$-jets in terms 
of the measured values $\tilde{X}_{1,2}$ and $\alpha_{1,2}^g $:
\begin{equation}\label{eqs:DDRSolution-withFNU}
\begin{split}
   X^q = \frac{\alpha_2^g \tilde{X}_1 - \alpha_1^g \tilde{X}_2}{\alpha_2^g - \alpha_1^g} , \,\,\,\,\,\,\,\,\,\,
   X^g = \frac{(1-\alpha_1^g) \tilde{X}_2 - (1-\alpha_2^g) \tilde{X}_1}{\alpha_2^g - \alpha_1^g}.
\end{split}
\end{equation}
  These characteristics refer to the combined jet sample.
  Decomposition of the characteristic $X$ of the combined jet sample into $q$- and $g$-jet 
components has the usual form:
\begin{equation}\label{eqs:combined-1}
  X \equiv \delta_1 X_1 + \delta_2 X_2 = \alpha^g X^g + (1 - \alpha^g) X^q,
\end{equation}
where $\delta_k = \frac{N_k}{N_1 + N_2}$ is fraction of jets of the $k$th jet sample 
in the combined one.
  The $g$-jet fraction in the combined jet sample is expressed in terms of $g$-jet fractions in subsamples:
\begin{equation}\label{eqs:combine-2}
   \alpha^g = \delta_1 \alpha_1^g + \delta_2 \alpha_2^g.
\end{equation}

\label{sec:Ddistrib}
\section*{\bf MEASUREMENT OF QUARK AND GLUON $D$-DISTRIBUTIONS }
  
  To measure the fraction of $g$-jets in $k$th jet sample by the fitting procedure~(\ref{eqs:QGLFitFunction}), 
"nonuniversal" $D$-distributions of $q/g$-jets are used, which are the distributions found
for $k$th jet sample, $ H^{q/g\,MC}_k(D)$.
  Via combining the two jet samples, one can obtain the “universal” $D$-distributions 
of $q/g$-jets, $H^{q/g\,MC}(D)$, defined by second equation in Eqs.~(\ref{eqs:JFNU-universal}).
  Eqs.~(\ref{eqs:DDRSolution-withFNU}) allow to obtain the "universal" $D$-distributions 
of $q/g$-jets, $H^{q/g\,DAT}(D)$, from the data.
  As model distributions $H^{q/g\,MC}_k(D)$ were used to obtain $\alpha_k^g$, 
the calculation of $H^{q/g\,DAT}(D)$ by the Eqs.~(\ref{eqs:DDRSolution-withFNU}) 
can be considered as a data-motivated correction of model "universal" 
distributions $H^{q/g\,MC}(D)$. 
  From Eqs.~(\ref{eqs:DDRSolution-withFNU}):
\begin{equation}\label{eqs:DDRSolution-withFNU-QGL}
\begin{split}
   H^{q\,DAT}(D) &= \frac{\alpha_2^g \tilde{H}_1(D) - \alpha_1^g \tilde{H}_2(D)}{\alpha_2^g - \alpha_1^g} , \\
   H^{g\,DAT}(D) &= \frac{(1-\alpha_1^g) \tilde{H}_2(D) - (1-\alpha_2^g) \tilde{H}_1(D)}{\alpha_2^g - \alpha_1^g}.
\end{split}
\end{equation}
where the quantities $\tilde{H}_{1,2}(D)$ are defined by 
Eqs.~(\ref{eqs:QGL-FIT-withFNU-tilde-where1}) -~(\ref{eqs:QGL-FIT-withFNU-tilde-where2}).

  For jet sample, which is a combination of two jet samples, 
the $D$-distribution has the form~(\ref{eqs:combined-1}):
\begin{equation}\label{eqs:combined-3}
    H^{DAT}(D) \equiv \delta_1 H_1^{DAT}(D) + \delta_2 H_1^{DAT}(D) = \alpha^g H^{g\, DAT}(D) + (1 - \alpha^g) H^{q\, DAT}(D).
\end{equation}
  This is an analytical expression for the $D$-distribution of the combined jet sample.
  From it, one can conclude that the $g$-jet fraction in the combined jet sample 
can be found by fitting using corrected $D$-distributions of $q/g$-jets:
\begin{equation}\label{eqs:combined-4}
     H^{DAT}(D) \sim \alpha^g H^{g\, DAT}(D) + (1 - \alpha^g) H^{q\, DAT}(D).
\end{equation}
  On the other hand, according to Eq.~(\ref{eqs:QGLFitFunction}), the $g$-jet fraction 
in the combined jet sample can be found by fitting using generator $q/g$-jet $D$-distributions:
\begin{equation}\label{eqs:combined-5}
     H^{DAT}(D) \sim \alpha^g H^{g\,MC}(D) + (1 - \alpha^g) H^{q\,MC}(D).
\end{equation}
  Although fitting functions in  Eqs.~(\ref{eqs:combined-4}) and~(\ref{eqs:combined-5}) 
are different, mean g-jets fractions in both cases match each other within uncertainties.

   In~\cite{QGL8TEV:2013, QGL13TEV:2017} "data driven reshaping" (DDR) procedure 
was proposed to correct the model $q/g$-jet $D$-distributions, $H^{q/g\,MC}(D)$,
using data. 
  DDR also uses two jet samples.
  In contrast to the procedure described above with the final equations~(\ref{eqs:DDRSolution-withFNU-QGL}), 
the original equations (Eqs.~(\ref{eqs:DDR})) were written in method DDR
for unnormalized histograms.
  In this form, the $g$-jet fractions are hidden.
  However, these equations implicitly contain the generator values $\alpha_k^{g, MC}$.
  
  To show this, write down the equations that are used in~\cite{QGL13TEV:2017} for the DDR procedure.
  Consider here a general case, which is presented in~\cite{QGL13TEV:2017}, 
in which jets with unidentified flavour are separated from $q$-jets.
  In this case, the original non-normalized $D$-distributions of jets 
in each MC jet sample have the form (see~\cite{QGL13TEV:2017}, p.~14):
\begin{equation}\label{eqs:MCdistribs}
\begin{split}
  N_1^{MC}(D) = N_1^{g\,MC}(D) +  N_1^{q\,MC}(D) + N_1^{x\,MC}(D), \\
  N_2^{MC}(D) = N_2^{g\,MC}(D) +  N_2^{q\,MC}(D) + N_2^{x\,MC}(D).
\end{split}
\end{equation}
  The numbers of jets in jet samples are equal: $N = N_{k}^{MC\, tot} = \sum_D N_k^{MC}(D)$.
  Instead of the index notation of $D$-discriminat bin $i$, which is used in~\cite{QGL13TEV:2017}, 
the functional notation is used here: $X(D) \equiv X_i$.
  For data, two equations are written for the same number of jets $N$ 
for two jet samples with unknown weights $w^q(D)$ and $w^g(D)$,
which are in~\cite{QGL13TEV:2017} are selected independent 
of jet sample, i.e. "universal" (in the terminology of this work):
\begin{equation}\label{eqs:DATdistribs}
\begin{split}
  N_1^{DAT}(D) = w^g(D) N_1^{g\,MC}(D) + w^q(D) N_1^{q\,MC}(D) + N_1^{x\,MC}(D), \\
  N_2^{DAT}(D) = w^g(D) N_2^{g\,MC}(D) + w^q(D) N_2^{q\,MC}(D) + N_2^{x\,MC}(D).
\end{split}
\end{equation}
  The weights correct the content of MC distributions in the $D$-bin 
so that the total $D$-distributions in the right sides of Eqs.~(\ref{eqs:DATdistribs}) 
reproduces the data $D$-distributions in the left sides.
  From Eqs.~(\ref{eqs:DATdistribs}) follows a system of equations with 
normalized distributions that explicitly contain the $g$-jet fractions:
\begin{equation}\label{eqs:DATdistribsNorm}
\begin{split}
  H_1^{DAT}(D) = w^q(D) \alpha_1^{g\,MC} H^{g\,MC}(D) + w^g(D) \alpha_1^{q\,MC} H^{q\,MC}(D) + \alpha_1^{x\,MC} H^{x\,MC}(D), \\
  H_2^{DAT}(D) = w^q(D) \alpha_2^{g\,MC} H^{g\,MC}(D) + w^g(D) \alpha_2^{q\,MC} H^{q\,MC}(D) + \alpha_2^{x\,MC} H^{x\,MC}(D).
\end{split}
\end{equation}
    Here the $f$-jet fraction is determined by the ratio $\alpha_k^{f\, MC} = N_k^{f\, MC\, tot}/N$, 
where $N_k^{f\, MC\, tot}$ is the number of $f$-jets in the $k$th MC jet sample
($N_k^{g\,MC\,tot} + N_k^{q\,MC\,tot} + N_k^{x\,MC\,tot} = N,\,\,\alpha_k^{g\,MC} + \alpha_k^{q\,MC} + \alpha_k^{x\,MC} = 1$).
  The JFNU correction is not taken into account in Eqs.~(\ref{eqs:DATdistribsNorm}), 
i.e. normalized $D$-distributions of $f$-jets are considered independent of the jet sample index $k$.
  The fact that JFNU correction is not taken into account 
justifies the assumption used in~\cite{QGL13TEV:2017} 
that the weights $w^f(D)$ in the Eqs.~(\ref{eqs:DATdistribs}) 
do not depend on index of jet sample $k$.
  
  System of Eqs.~(\ref{eqs:DATdistribsNorm}) contains decompositions 
of the $D$-distributions for data jet samples into $q$- and $g$-jet $D$-distributions 
that are normalized  by the value:
\begin{equation}\label{eqs:normFactor}
   S^f \equiv \sum_D w^f(D) H^{f\,MC}(D).
\end{equation}
  One can introduce corrected by DDR procedure 
normalized $D$-distributions of $f$-jets, $H^{g\,DAT\prime}(D)$, 
and corrected fractions of $f$-jets, $\alpha_1^{g\prime}$:
\begin{equation}\label{eqs:DATdistribsNormPrime}
\begin{split}
  H_1^{DAT}(D) = \alpha_1^{g\prime} H^{g\,DAT\prime}(D) + \alpha_1^{q\prime} H^{q\,DAT\prime}(D) + \alpha_1^{x\,MC} H^{x\,MC}(D), \\
  H_2^{DAT}(D) = \alpha_2^{g\prime} H^{g\,DAT\prime}(D) + \alpha_2^{q\prime} H^{q\,DAT\prime}(D) + \alpha_1^{x\,MC} H^{x\,MC}(D),
\end{split}
\end{equation}
where
$\alpha_k^{f\prime} \equiv S^f \alpha_k^f$, $H^{f\prime}(D) \equiv w^f(D) H^{f\,MC}(D)/S^f$.
  It follows from Eqs.~(\ref{eqs:DATdistribsNormPrime}) that 
$\alpha_k^{g\prime} + \alpha_k^{q\prime} + \alpha_k^x = 1$.
  Thus, to obtain $q(g)$-jet fraction in $k$th data jet samples it is necessary
to multiply MC $q(g)$-jet fraction by the factors $S^q$ and $S^g$ respectively.
  The same rusults can be obtained by fit~(\ref{eqs:QGLFitFunction}) with corrected distributions
$H^{f\,DAT\prime}(D)$.
  
  Moreover, it follows from Eqs.~(\ref{eqs:DATdistribsNormPrime}) that $S^g=S^q=1$.
  To show this, one can sum up the Eqs.~(\ref{eqs:DATdistribsNormPrime}) by bins 
and take into account the normalization of histograms:
\begin{equation}\label{eqs:DATdistribsNormPrimeSum}
\begin{split}
  \alpha_1^{g\,MC} S^g + \alpha_1^{q\,MC} S^q = \alpha_1^{g\,MC} + \alpha_1^{q\,MC}, \\
  \alpha_2^{g\,MC} S^g + \alpha_2^{q\,MC} S^q = \alpha_2^{g\,MC} + \alpha_2^{q\,MC}.
\end{split}
\end{equation}
  The only solution of this system is $S^g=S^q=1$ if the determinant is not zero
$\Delta = \alpha_1^{g\,MC}\alpha_2^{q\,MC} - \alpha_1^{q\,MC}\alpha_2^{g\,MC} \ne 0 $.
  
  In the case $q=\{u,d,s,c,b,x\}$  (i.e. $\alpha_1^{x\,MC}=0$ and
$\alpha_k^{q\,MC} =  1 - \alpha_k^{g\,MC}$) condition $\Delta\ne 0$ 
is condition of inequality of $g$-jet fractions between two jet samples 
($\alpha_1^g \ne \alpha_2^g$) and system of Eqs.~(\ref{eqs:DATdistribsNormPrime})
take the form:
\begin{equation}\label{eqs:DATdistribsNormPrimeWox-1}
\begin{split}
  H_1^{DAT}(D) = \alpha_1^{g\,MC} H^{g\,DAT\prime}(D) + \alpha_1^{q\,MC} H^{q\,DAT\prime}(D), \\
  H_2^{DAT}(D) = \alpha_2^{g\,MC} H^{g\,DAT\prime}(D) + \alpha_2^{q\,MC} H^{q\,DAT\prime}(D),
\end{split}
\end{equation}

  The solution of system~(\ref{eqs:DATdistribsNormPrimeWox-1}) relative to 
unknown distributions $H^{q/g\,DAT\prime}(D)$ is equivalent 
to solution of the system of equations for weights~(\ref{eqs:DATdistribs}) 
and, after that, calculation of $H^{q/g\,DAT\prime}(D) = w^{q/g}(D) H^{q/g\,MC}(D)$.
  The solution of system~(\ref{eqs:DATdistribsNormPrimeWox-1}) 
has the form~(\ref{eqs:DDRSolution-withFNU-QGL}),
where it is necessary to replace $\alpha_k^g$ 
($g$-jet fraction in  $k$th data jet sample) with $\alpha_k^{g\, MC}$ ($g$-jet fraction in $k$th 
MC jet sample) and neglect the JFNU correction.
  In real analysis, after applying the DDR procedure, 
one can see some small deviation of the $g$-jet fractions between data and MC 
within the JFNU correction.


  Eqs.~(\ref{eqs:DDRSolution-withFNU-QGL}) correct the DDR procedure by replacing
MC $g$-jet fractions $\alpha_k^{g,\,MC}$ with the measured ones, $\alpha_k^g$.
  In addition, the Eqs.~(\ref{eqs:DDRSolution-withFNU-QGL}) contain JFNU correction, 
  which describes universally  the differences of $q/g$-jet characteristics 
  in different jet samples, and which is important if the fractions of $g$-jets 
  are measured with good accuracy.
  
  Thus, the Eqs.~(\ref{eqs:DDRSolution-withFNU-QGL})
correct the DDR procedure by replacing the generator $g$-jet fractions $\alpha_k^{g, MC}$ 
with the measured ones, $\alpha_k^g$.
  Inclusion of the JFNU correction also distinguishes the procedure 
for finding $H^{q/g\, DAT}(D)$ by Eqs.~(\ref{eqs:DDRSolution-withFNU-QGL}) 
from the DDR procedure.

\label{sec:summary}
\section*{\bf CONCLUSION }

  The article presents a technique for measuring characteristics 
of quark and gluon jets, using measurement of gluon 
jet fractions in two control jet samples.
  
  The most effective way of measuring gluon jet fraction is 
fitting measured jet $D$-distribution ($D$ is quark/gluon 
jet discriminator) with a linear combination of 
quark and gluon jets $D$-distributions with gluon jet fraction 
as a fitting parameter.
  Quark and gluon jets $D$-distributions are constructed 
for each jet sample separately using a generator model 
and with full simulation of detector response.
  The model defines quark/gluon jet physics parameters,
which is used to construct jet $D$-value  
and a rule for identifying the jet type  – a rule to match 
parton initiating the jet and jet at hadron level.

  Two jet samples characteristics divergence is defined mainly 
by the quark/gluon jet fractions difference between the jet samples.
  This is due to the fact that the properties of quark and gluon jets 
differ significantly.
  For precision measurement of quark and gluon  jet characteristics, 
one should also consider the following: there is some 
difference in quark and gluon jet characteristics between 
the jet samples.
  This difference arises from jet kinematics and/or 
from jet physical environment, which can affect the operation 
of jet finder algorithm.

  A universal method for taking into account the differences in 
the quark/gluon jet characteristics between the two jet samples 
is proposed in the article.
  The method introduces a notion of a "measure of non-universality" 
of quark/gluon jet characteristic, which is defined as a difference
of characteristic between the two jet samples, 
and a notion of "universal" (averaged) quark/gluon jet characteristic, 
which is a characteristic of jet sample that combines 
the two jet samples under study.
  It is shown that the "universal" characteristic of 
quark/gluon jets are expressed via the measured characteristics 
for the two jet samples with some correction.
  The correction contains "measures of non-universality" 
of quark and gluon jet characteristics 
(determined in the Monte Carlo model), 
measured gluon jet fractions, 
and ratio of number of jets in the two jet samples.
 
   The quark-gluon discriminator $D$, which is used 
to measure gluon jet fraction, is considered as an example 
of the measured jet sample characteristic.
   Following general formalism, the quark and gluon 
"universal" $D$-distributions, related to the combined 
jet sample, can be found from the data.
  These distributions can be compared with the original 
model quark and gluon $D$-distributions, which were used 
to measure gluon jet fractions.
  It will allow to determine how well the model 
of quark and gluon jet formation agrees with 
the real process.



\end{document}